\documentclass[5p]{elsarticle}

\usepackage{amssymb}

\journal{TIPP09 Proceedings in NIMA}

\begin{document}

\begin{frontmatter}



\title{The ATLAS beam pick-up based timing system}

%
%
%
%
%
\author[First]{C.\, Ohm\corref{cor1}}
\ead{christian.ohm@cern.ch}
\author[Second]{T.\,Pauly}
\cortext[cor1]{Corresponding author. Tel.: +41-(0)79-771-2149; Office: +41-(0)22-767-1126.} 
%
%
%
\address[First]     {Stockholm University, Department of Physics, 106 91 Stockholm, Sweden}
\address[Second]     {European Organization for Nuclear Research, 1211 Gen\`eve, Switzerland}

%
%
%
%
%
\begin{abstract}

The ATLAS BPTX stations are comprised of electrostatic button pick-up detectors, located 175 m away along the beam pipe on both sides of ATLAS. The pick-ups are installed as a part of the LHC beam instrumentation and used by ATLAS for timing purposes. 

The usage of the BPTX signals in ATLAS is twofold: they are used both in the trigger system and for LHC beam monitoring. The BPTX signals are discriminated with a constant-fraction discriminator to provide a Level-1 trigger when a bunch passes through ATLAS. Furthermore, the BPTX detectors are used by a stand-alone monitoring system for the LHC bunches and timing signals. The BPTX monitoring system measures the phase between collisions and clock with a precision better than 100 ps in order to guarantee a stable phase relationship for optimal signal sampling in the subdetector front-end electronics. In addition to monitoring this phase, 
the properties of the individual bunches are measured and the structure of the beams is determined. 

On September 10, 2008, the first LHC beams reached the ATLAS experiment. During this period with beam, the ATLAS BPTX system was used extensively to time in the read-out of the sub-detectors. In this paper, we present the performance of the BPTX system and its measurements of the first LHC beams.
\end{abstract}

%
%
%
%
%
%
\begin{keyword}

ATLAS \sep
Beam monitoring \sep
Level-1 trigger \sep
BPTX \sep
LHC \sep
LHC timing signals



\end{keyword}

\end{frontmatter}


%
%
%
%
%
%

\section{Introduction}

The ATLAS experiment \cite{detectorpaper} at the Large Hadron Collider (LHC) \cite{lhcmachinepaper} must be synchronized to the collisions to ensure the quality of the event data recorded by its sub-detectors. In order to facilitate this, the LHC provides beam related timing signals to the experiments via optical fibers that are several kilometers long \cite{ttc}. The phase of these clock signals can drift, e.g. due to temperature fluctuations, causing front-end electronics to sample at non-optimal working point. 
On both sides of ATLAS, 175\,m upstream from the interaction point, beam pick-up detectors are installed along the LHC beam pipe. This paper describes how these detectors are used
   \begin{itemize}
   \item \vspace{-0.6em} to monitor the phase between the collisions and the LHC clock signals that drive the ATLAS electronics
   \item \vspace{-0.6em} to monitor the structure and uniformity of the LHC beams
   \item \vspace{-0.6em} as input to the trigger system
   \end{itemize}


\section{The BPTX detectors}
\begin{figure}[t!]
  \begin{center}
  \includegraphics[width=0.4\textwidth]{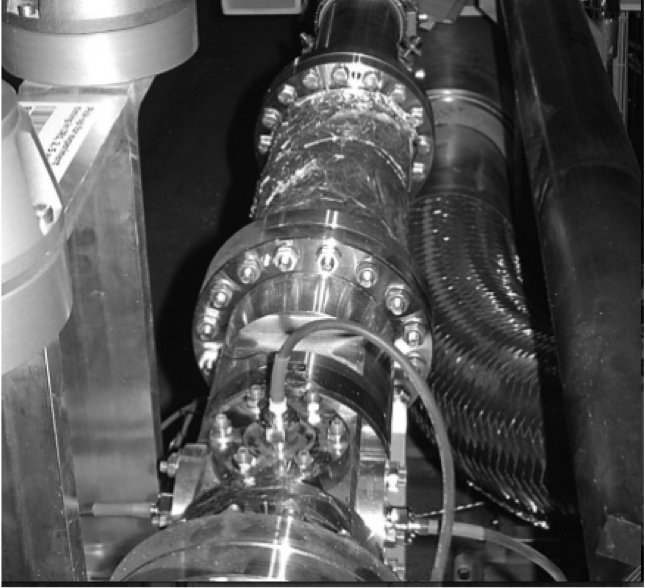}
  \caption{A photograph of one of the two ATLAS BPTX stations.}
  \vspace{-1em}
  \label{fig:bptxstation}
  \end{center}
\end{figure}

The BPTX stations are beam position monitors provided by the LHC machine, but operated by experiments for timing purposes. They are comprised of four electrostatic button pick-up detectors, arranged symmetrically in the transverse plane around the LHC beam pipe. Since the signal from a passing charge distribution is linearly proportional to distance to first order, the signals from all four pick-ups are summed to cancel out potential beam position deviations. The resulting signal is then transmitted to the underground counting room \emph{USA15} via a 220\,m low-loss cable. Figure~\ref{fig:bptxstation} shows the installed BPTX station for beam 2 on the C-side of ATLAS. At the bottom of the photograph, the cables from the four button pick-ups are visible.

\section{Usage of the beam pick-up signals}
\begin{figure}[h!]
	\begin{center}
	\includegraphics[width=0.4\textwidth]{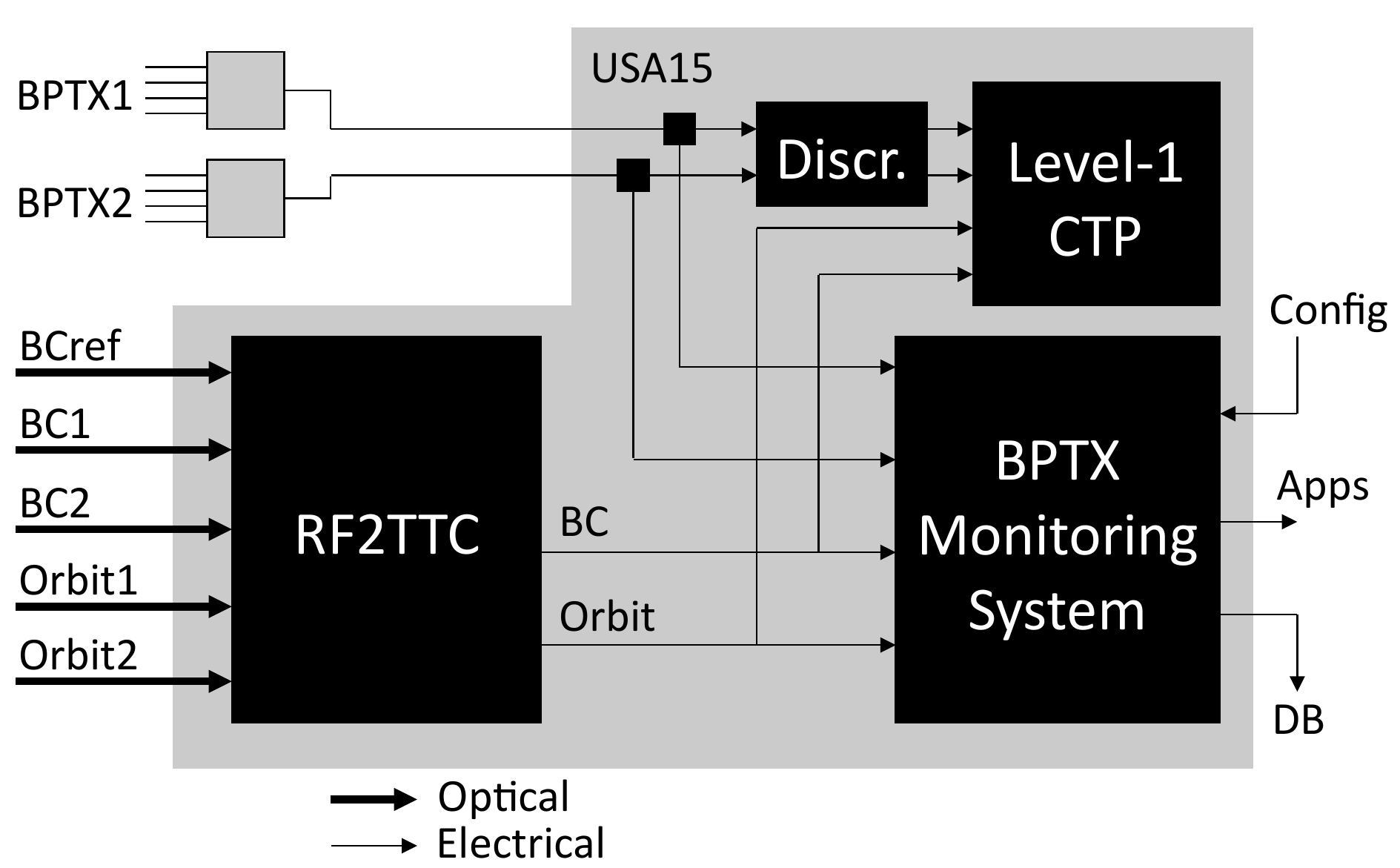}
	\caption{Diagram showing the BPTX system and how it interacts with the related systems in ATLAS.}
	\label{fig:bptxcontextdiagram}
	\end{center}
\end{figure}

The BPTX signals are used for two separate purposes within ATLAS, by the trigger system and by a monitoring system for the LHC beams and timing signals. Figure~\ref{fig:bptxcontextdiagram} shows the BPTX system and how it interacts with the related systems \cite{timing-in,masterohm}. The optical timing signals from the LHC arrive in the underground counting room to a receiver module, the \verb=RF2TTC=. This module converts the optical signals to \emph{TTC}\footnote{TTC is the standard hardware system used across the LHC experiments for distribution of fast Timing, Trigger and Control signals} signals and can also manipulate their phase, duration etc. if needed. The electrical signals are then transmitted to the ATLAS sub-detectors via the \emph{Central Trigger Processor} (CTP) of the Level-1 trigger system and to the BPTX monitoring system.

\subsection{Level-1 Trigger}
The ATLAS trigger system is designed in three levels, each level sequentially refining the selection of events to be saved for further offine analysis. The Level-1 trigger is implemented in custom electronics and performs a first selection of events within 2.5 $\mu$s, based primarily on reduced-granularity data from the calorimeters and the muon spectrometer. The selected events are processed further by the \emph{High Level Trigger} system which is implemented in software. The signals from the BPTX stations are discriminated with a constant-fraction discriminator to provide ATLAS with an accurate and reliable timing reference in the form of a standard NIM pulse. This pulse is fed into the Level-1 \emph{Central Trigger Processor} where it serves as a trigger condition indicating a bunch passing through ATLAS. 

\subsection{Monitoring of the LHC beams and timing signals}
Furthermore, the BPTX detectors are used by a stand-alone monitoring system for the LHC bunches and timing signals. The BPTX and LHC timing signals are digitized by a deep-memory, high sampling rate (5\,GHz) oscilloscope and transferred to a computer running Linux for analysis. The features of the scope enables capturing a full LHC turn in one acquisition while retaining enough detail to get about 5 measurement points on the sharp falling edge of each BPTX pulse (see e.g. Figure~\ref{fig:firstbunch}). Since most of the high-frequency content of the BPTX signals is attenuated by the long transmission line, the frequency spectrum of the signals arriving in ATLAS peaks around 400 MHz, making an analog bandwidth of 600 MHz sufficient for the oscilloscope used for digitization. By making fits to the identified bunch pulses and clock edges, the BPTX monitoring system measures the phase between each bunch and the clock signal with high accuracy. Monitoring these quantites is crucial to guarantee a stable phase relationship for optimal signal sampling in the subdetector front-end electronics. In addition to monitoring this phase, the intensity and longitudinal length of the individual bunches are measured and the structure of the beams is determined. Using the BPTX monitoring applications, the shifter in the control room can verify that the timing signals are synchronized to the collisions, and also look for so-called \emph{satellite bunches}, out-of-time bunches that would cause off-center collisions in ATLAS.

The monitoring system is running independently from the ATLAS online data acquisition infrastructure, enabling monitoring of the LHC machine in the control room even when ATLAS is not taking data. Summary data from the BPTX monitoring system, e.g. mean bunch intensity and phase, is published to the ATLAS \emph{Detector Control System}\cite{dcs} and ultimately saved to the conditions database.

\section{Results from the first LHC beams}
\subsection{The first proton bunches in ATLAS}
\begin{figure}[h!]
	\begin{center}
	\includegraphics[width=\linewidth]{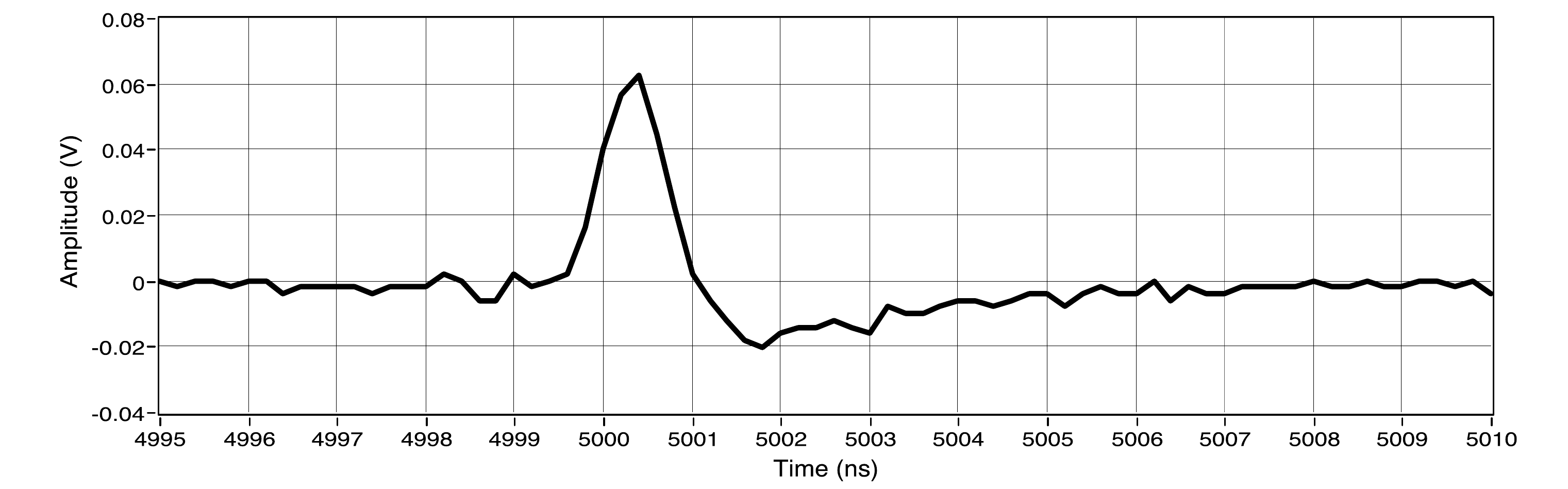}
	\caption{The first LHC bunch on its way to ATLAS.}
	\label{fig:firstbunch}
	\end{center}
\end{figure}

On September 10, 2008, the first LHC proton bunch reached ATLAS. Figure \ref{fig:firstbunch} shows the pulse recorded by the BPTX monitoring system. 

\begin{figure}[h!!]
	\begin{center}
	\includegraphics[width=\linewidth]{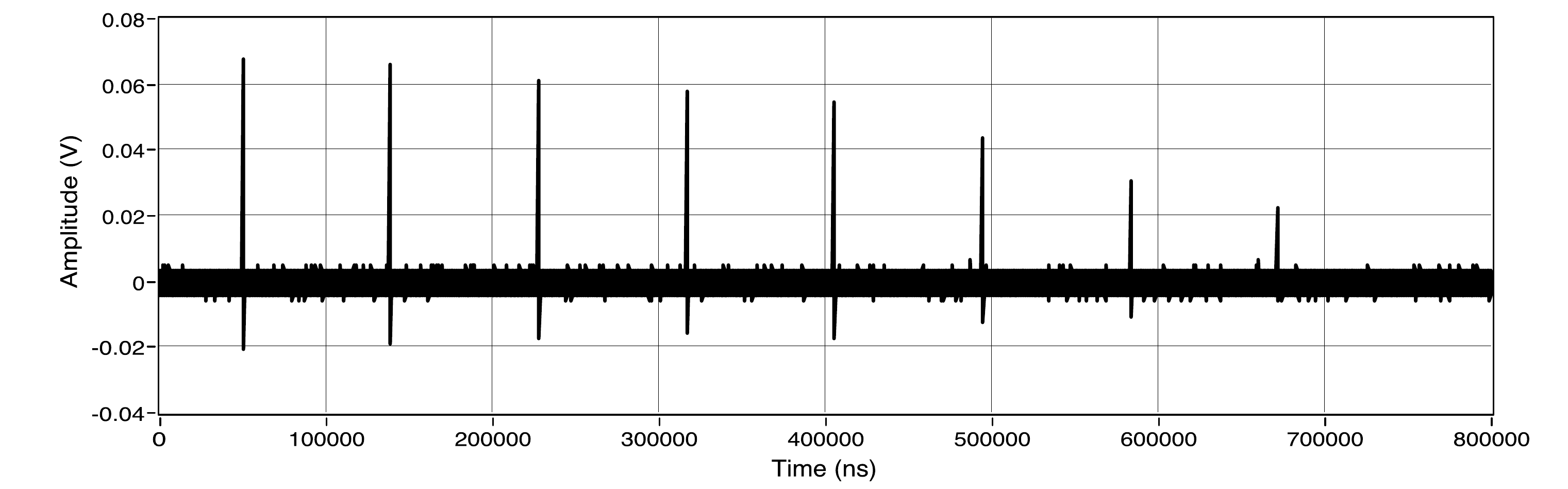}
	\caption{A bunch passing ATLAS in 8 consecutive turns.}
	\label{fig:8turns}
	\end{center}
\end{figure}

A few hours later, a bunch was successfully circulated 8 turns around the accelerator and seen by ATLAS as depicted in Figure~\ref{fig:8turns}. The pulses are separated by 89\,$\mu$s, corresponding to the time it takes for an LHC bunch to circulate around the 27\,km long ring. The pulse amplitude, which is proportional to the bunch intensity, is degrading from turn to turn, which is consistent with the beam loss and debunching expected for a beam not yet captured by the LHC RF system.

\subsection{Monitoring of a longer LHC run}
\begin{figure}[h!]
	\begin{center}
	\includegraphics[width=0.4\textwidth]{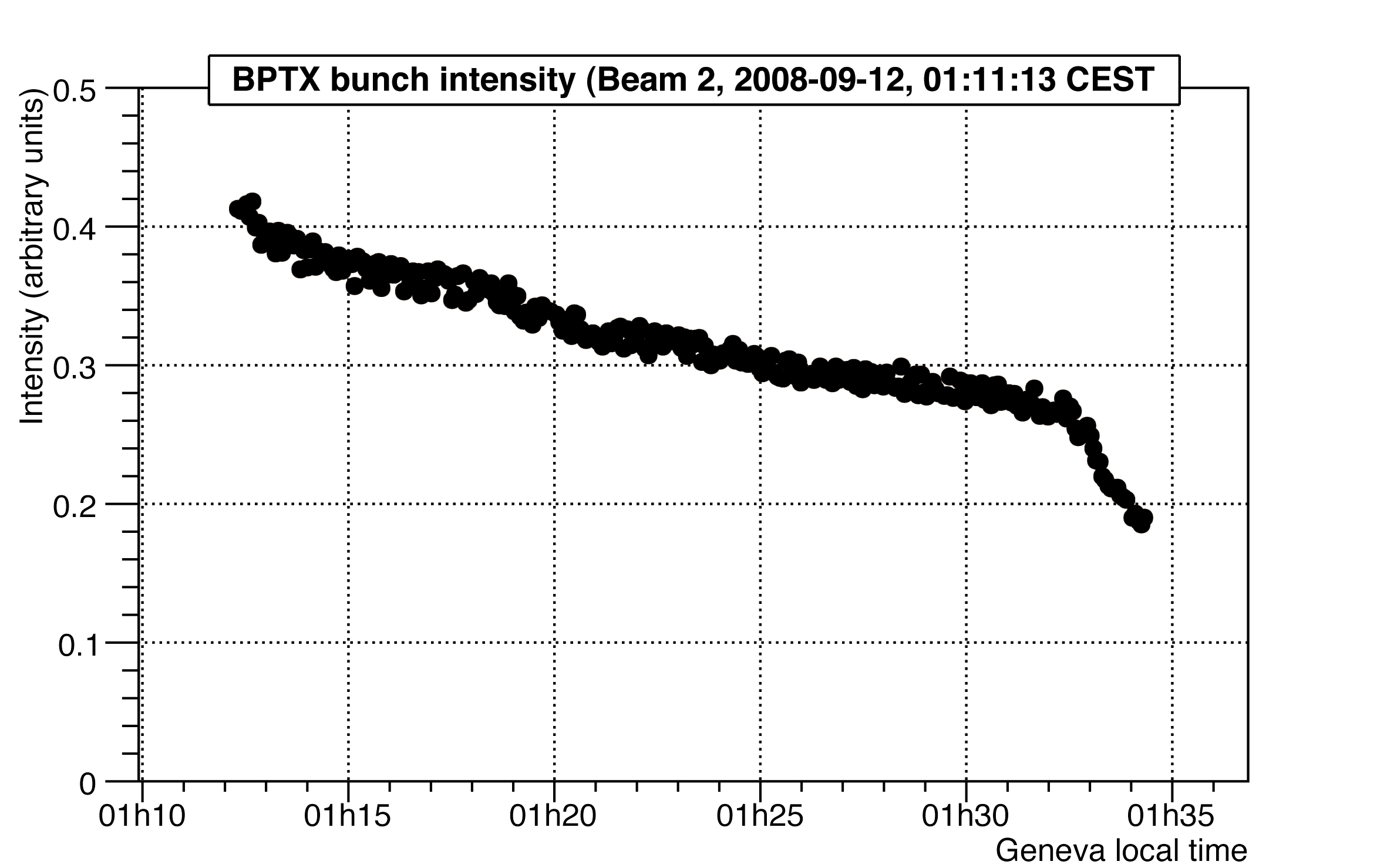}
	\caption{Intensity measured by the BPTX monitoring system during 20 minutes of circulating beam.}
	\label{fig:longfill}
	\end{center}
\end{figure}

Around 1 AM on September 12, 2008, a single bunch was circulated around the LHC for about 20 minutes after being captured by the RF system. The BPTX monitoring system measured the intensity during this period, and the resulting plot is shown in Figure~\ref{fig:longfill}. It should be noted that this is a relative but not yet normalized intensity measurement. The scattering of the data points suggests that the precision is around 10\%.

\begin{figure}[h!]
	\begin{center}
	\includegraphics[width=0.4\textwidth]{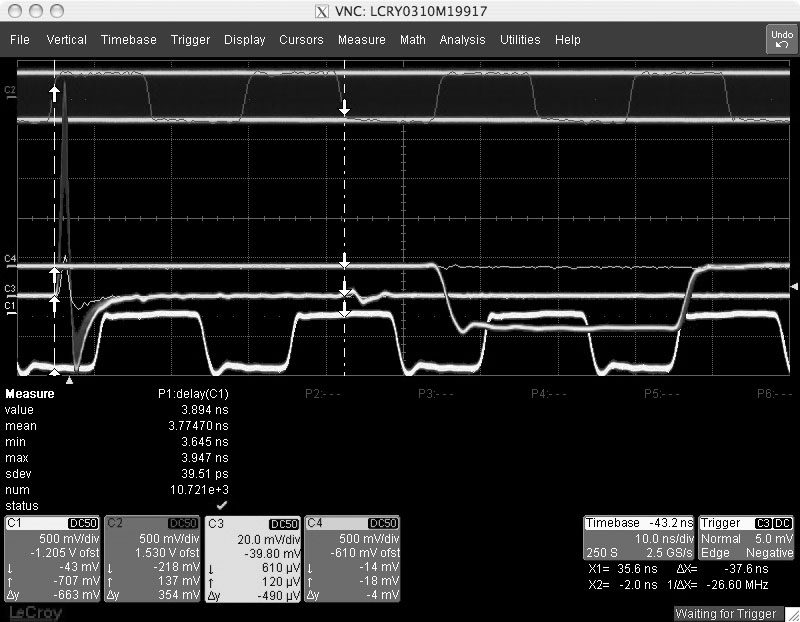}
	\caption{Oscilloscope traces from 20 minutes of circulating beam with persistency.}
	\label{fig:persistency}
	\end{center}
\end{figure}

Figure~\ref{fig:persistency} shows an oscilloscope picture recorded in persistency mode during the same time period. The falling edge of the analog BPTX signal for beam 2 (the scope channel with bipolar pulses to the left) is used as scope trigger and can be seen together with the discriminated BPTX signal used as Level-1 trigger input (with longer NIM pulse to the right). The clock related to beam 2 (bottom) is stable within an RMS of 40 picoseconds with respect to the beam, indicating RF capture. The reference clock signal (top), corresponding to the bunch frequency at a higher energy, has a different frequency.

\section{Conclusions}
In the first period of beam in the LHC, the BPTX system was used extensively as a trigger to time in the read-out windows of the sub-detectors of the ATLAS experiment. The BPTX monitoring system was able to record the very first LHC bunch approaching ATLAS, and provided detailed information about the beams during these first days of data taking.

\section*{Acknowledgments}

The authors would like to thank the ATLAS Collaboration and its Level-1 Central Trigger group in which most of this work was carried out. We would also like to express our deepest gratitude to the LHC community, both for the support and for providing the BPTX detectors.

%
%
%
%
%
%

\end{document}